\documentclass[12pt]{article}
\usepackage{epsf}
\def\be{\begin{equation}}
\def\ee{\end{equation}}
\def\bea{\begin{eqnarray}}
\def\eea{\end{eqnarray}}

\begin{document}
\begin{titlepage}
\begin{center}
{\Large \bf William I. Fine Theoretical Physics Institute \\
University of Minnesota \\}
\end{center}
\vspace{0.2in}
\begin{flushright}
FTPI-MINN-13/32 \\
UMN-TH-3303/13 \\
September 2013 \\
\end{flushright}
\vspace{0.3in}
\begin{center}
{\Large \bf $Y(4260)$ and $Y(4360)$ as  mixed hadrocharmonium
\\}
\vspace{0.2in}
{\bf Xin Li$^a$  and M.B. Voloshin$^{a,b,c}$  \\ }
$^a$School of Physics and Astronomy, University of Minnesota, Minneapolis, MN 55455, USA \\
$^b$William I. Fine Theoretical Physics Institute, University of
Minnesota,\\ Minneapolis, MN 55455, USA \\
$^c$Institute of Theoretical and Experimental Physics, Moscow, 117218, Russia
\\[0.2in]

\end{center}

\vspace{0.2in}

\begin{abstract}
Recent BESIII data indicate a significant rate of the process $e^+e^- \to h_c \pi^+ \pi^-$  at the $Y(4260)$ and $Y(4360)$ resonances, implying a substantial breaking of the heavy quark spin symmetry. We consider these resonances within the picture of hadrocharmonium, i.e. of (relatively) compact charmonium embedded in a light quark mesonic excitation. We suggest that the resonances $Y(4260)$ and $Y(4360)$ are a mixture, with mixing close to maximal, of two states of hadrochamonium, one containing a spin-triplet $c \bar c$ pair and the other containing a spin-singlet heavy quark pair. We argue that this model is in a reasonable agreement with the available data and produces distinctive and verifiable predictions for the energy dependence of the production rate in $e^+e^-$ annihilation of the final states $J/\psi \pi \pi$, $\psi' \pi \pi$ and $h_c \pi \pi$, including the pattern of interference between the two resonances. 
\end{abstract}
\end{titlepage}

The charmoniumlike resonances $Y(4260)$ and $Y(4360)$ in $e^+e^-$ annihilation present a considerable challenge for interpretation of their internal structure due to their unusual decay properties. Namely these resonances were mostly observed through their pionic transitions to either $J/\psi$ or $\psi'$ charmonium states: $Y(4260) \to J/\psi \pi \pi$~\cite{babar26,cleo26,belle26,babar26:12} and $Y(4360) \to \psi' \pi \pi$~\cite{babar36,belle36}. The most surprising feature of these resonances is that, unlike for other known states above the open charm threshold, e.g. $\psi(3770)$, or $\psi(4040)$, the decays to final states containing pairs of charmed mesons are not dominant. Several models for the structure of $Y(4260)$ have been discussed in the literature: a $c \bar c g$ hybrid~\cite{hyb26}, a $c s \bar c \bar s$ tetraquark~\cite{tetra26}, hadrocharmonium\cite{hc,dgv}, and, more lately as an $S$-wave molecular system containing an excited $D_1(2420)$ meson and a $D$ meson~\cite{molec26}. Most recently the new results from the BESIII experiment have added to the intrigue of the properties of the $Y(4260)$ and $Y(4360)$ resonances and may in fact hold a clue to understanding the structure of these states. Namely, in addition to the observation of isovector peaks $Z_c^\pm(3900)$~\cite{bes39} and $Z_c^\pm(4025)$~\cite{bes40} in the decays $Y(4260) \to Z_c(3900) \pi$ and $Y(4260) \to Z_c(4025) \pi$ the BESIII collaboration reported~\cite{beslp} an observation of  production of the final state $h_c \pi^+ \pi^-$ at both $\sqrt{s}= 4.26\,$GeV [$\sigma(e^+e^- \to h_c \pi^+ \pi^-) = 41.0 \pm 2.8 \pm 4.7\,$pb] and $\sqrt{s} = 4.36\,$GeV [$\sigma(e^+e^- \to h_c \pi^+ \pi^-) = 52.3 \pm 3.7 \pm 9.2\,$pb] , with a yield comparable to that of e.g 
$J/\psi \pi^+ \pi^-$ at the peak of $Y(4260)$: $\sigma(e^+e^- \to J/\psi \pi^+ \pi^-) = 62.9 \pm 1.9 \pm 3.7\,$pb. The latter behavior clearly implies a significant breaking of the heavy quark spin symmetry, which, if unbroken, would forbid the transitions to the spin singlet charmonium $1^1P_1$ state $h_c$. It may appear at first that this behavior is reminiscent of the known production of the $h_b(1P)$ and $h_b(2P)$~\cite{belleh} bottomonium spin singlet states in the two-pion transitions from $\Upsilon(10890)$. In the bottomonium case this apparent breaking of the heavy quark symmetry is entirely associated with the $Z_b(10610)$ and $Z_b(10650)$ isovector resonances~\cite{bellez} and the observed properties of these transitions are in agreement with the molecular picture~\cite{bgmmv} for the $Z_b$ resonances. The data however indicate that for the charmoniumlike resonances the dominant contribution to the transitions to $h_c \pi \pi$ is continually spread over the phase space, rather than being associated with an intermediate $Z_c$ resonance. Therefore one has to explain these transitions either by a heavy quark symmetry breaking within the $Y(4260)$ and $Y(4360)$ resonances, or in the mechanism for their decay. Furthermore, these two closely spaced resonances appear to display very similar properties, which may hint at a common structure of the two states.

In fact the splitting of the discussed resonances by about 100\,MeV can be compared with the characteristic scale of the heavy quark spin symmetry breaking in the charm sector for which a representative value is the mass splitting between $D^*$ and $D$ mesons of about 140\,MeV. In this paper we suggest and explore the possibility that the resonances $Y(4260)$ and $Y(4360)$ form a pair of mixed states containing both a spin-triplet and a spin-singlet $c \bar c$ pair. Clearly, such mixing is possible only in the presence of other degrees of freedom, i.e. of the light quarks and/or gluons, and is generally possible in either of the discussed models of hybrid or four quark systems.   A description of the properties of $Y(4260)$ and $Y(4360)$ based on molecular structure involving orbitally excited charmed mesons encounters certain difficulties~\cite{xlv}, and we rather discuss the mixing within the hadrocharmonium model~\cite{hc}, where a relatively compact colorless $c \bar c$ pair is embedded in a mesonic excitation of light quarks. 

In the hadrocharmonium model one can naturally expect a mixing between an embedded $^3S_1$ charmonium state and a $^1P_1$ state. Indeed, in terms of the multipole expansion in QCD~\cite{gottfried,mv79}, the leading interaction depending on the spin of the heavy quarks is the chromomagnetic dipole (M1), described by the Hamiltonian
\be
H_{M1}=-{1 \over 4 m_c} \, \xi^a \, (\vec \Delta \cdot \vec B^a)~,
\label{hm1}
\ee
where $\vec B^a$ is the chromomagnetic field, $\xi^a = t_c^a - t_{\bar c}^a$ is the difference of the color generators acting on the quark and antiquark, and $\vec \Delta = \vec \sigma_c - \vec \sigma_{\bar c}$ is a similar difference for the spin operators. The leading effect in transitions between colorless states of a nonrelativistic $c \bar c$ pair induced by this term arises through its interference with the chromoelectric dipole (E1) interaction
\be 
H_{E1}=-{1 \over 2} \, \xi^a \, (\vec r \cdot \vec E^a )~,
\label{he1}
\ee
where $\vec E^a$ is the chromoelectric field and $\vec r$ is the vector of the relative position between the quark and the antiquark. Clearly, the combined action of the terms (\ref{hm1}) and (\ref{he1}) changes the orbital angular momentum by one unit and the total spin of the pair by one unit, $\Delta L=1$ and $\Delta S=1$, and thus links a $^3S_1$ state of charmonium to the $^1P_1$.

In the present discussion we denote $\Psi_3$ the wave function of a hadrocharmonium state with the quantum numbers $J^{PC}=1^{--}$ containing a $^3S_1$  $c \bar c$ pair, and denote $\Psi_1$ that for the $J^{PC}=1^{--}$ state with an embedded $^1P_1$ $c \bar c$ pair~\footnote{Clearly the required overall quantum numbers $J^{PC}=1^{--}$ with a $c \bar c$ pair in the $^1P_1$ state can arise only in the hadrocharmonium system due to the contribution of the light degrees of freedom.}. We suggest that the observed $Y(4260)$ and $Y(4360)$ resonances arise as a result of mixing between these two states due to the spin dependent interaction:
\be
Y(4260) = \cos \theta \, \Psi_3 - \sin \theta \, \Psi_1~, ~~~~~~Y(4360) = \sin \theta \, \Psi_3 + \cos \theta \, \Psi_1
\label{mix}
\ee
with $\theta$ being the mixing angle.

Assuming that the mixing is the dominant source of the heavy quark spin symmetry breaking, the model of the mixed states described by Eq.(\ref{mix}) imples a distinctive pattern of production in the $e^+e^-$ annihilation of the final states $J/\psi \pi \pi$, $\psi' \pi \pi$ and $h_c \pi \pi$ in the energy region of the $Y(4260)$ and $Y(4360)$ resonances including the interference effects. Indeed, due to the heavy quark spin symmetry only the $\Psi_3$ state is produced by the electromagnetic current, and the decays to the final states with either the $J/\psi$ or $\psi'$ charmonium are due to the same $\Psi_3$. Thus the amplitudes for production of these final states can be written as
\be
A[e^+e^- \to J/\psi(\psi') \pi \pi] \propto \left ( \cos^2 \theta \, BW_1 + \sin^2 \theta \, BW_2 \right ) \, A[\Psi_3 \to J/\psi(\psi') \pi \pi]~,
\label{a3}
\ee
where $BW_1$ and $BW_2$ stand for the Breit-Wigner resonance factors for respectively $Y(4260)$ and $Y(4360)$, $BW(E) = (E-M+i \, \Gamma/2)^{-1}$ with $E$ being the c.m. energy. On the other hand, the production of the final state $h_c \pi \pi$ is exclusively due to the mixing with $\Psi_1$, so that the production amplitude reads as
\be
A[e^+e^- \to h_c \pi \pi] \propto \cos \theta \, \sin \theta \, \left ( BW_1 - BW_2 \right ) \, A[\Psi_1 \to h_c \pi \pi]~.
\label{a1}
\ee
The proportionality coefficients in Eqs. (\ref{a3}) and (\ref{a1}) depend on unknown couplings, so that these formulas can be used to describe the behavior of the yield in each channel in the resonance region, but not, say, the relative yield for different channels.

It can be noted that the relative sign between the two Breit-Wigner factors in Eqs.~(\ref{a3}) and (\ref{a1}) is uniquely determined by the inherent in the discussed hadrocharmonium model assumption that the structures $Y(4260)$ and $Y(4360)$ arise from the mixing of two states with definite total spin of the $c \bar c$ pair, i.e. $\Psi_3$ with $S_{c \bar c} =1$ and $\Psi_1$ with $S_{c \bar c} =0$. This implies that in $\Psi_3$ and $\Psi_1$ the heavy quark and antiquark are correlated with each other, rather than each having a strong correlation with the light constituents, which would be the case in a molecular, tetraquark, or hybrid picture. In the latter models individual states would be mixed in the total spin of the $c \bar c$ pair, so that the interference pattern between $Y(4260)$ and $Y(4360)$ in the discussed final channels would generally be different. 

The behavior of the amplitudes $A[\Psi_3 \to \psi' \pi \pi]$ and $A[\Psi_1 \to h_c \pi \pi]$ is in all likelihood somewhat different from that of $A[\Psi_3 \to J/\psi \pi \pi]$. Namely, the energy released in the pion pair in the former two processes is sufficiently low, and one can rely on the chiral low energy regime, where the amplitudes for these decays are bilinear in the energy or momentum of the two pions~\cite{bc,mv75}, and this behavior is in agreement with the reported~\cite{babar36,belle36} pion spectra in the transitions $Y(4360) \to \psi' \pi \pi$. Therefore, neglecting the pion mass, one can approximate the rate for these decays is being proportional to the seventh power of the energy release,
\be
\Gamma [\Psi_3 \to \psi' \pi \pi] \propto [E-M(\psi')]^7~, \Gamma [\Psi_1 \to h_c \pi \pi] \propto [E-M(h_c)]^7~.
\label{g31}
\ee
Clearly, the strong dependence on energy enhances the rates for the higher peak $Y(4360)$, and the effect is most significant for the emission of $\psi'$ where the available energy is the smallest. 
On the other hand, the energy release in the transitions to $J/\psi \pi \pi$ exceeds 1.1\,GeV, so that the low energy chiral limit is not applicable. Rather one would expect that these latter transitions are dominated by the $f_0(980)$ resonance in the dipion channel, which expectation is supported by the available data~\cite{babar26,cleo26,belle26}. Thus the decay can be approximated as a two-body process: $\Psi_3 \to J/\psi f_0$, so that there is very little phase space kinematical dependence over the energy range of the $Y(4260)$ and $Y(4360)$ resonances. 

The approximation in Eq.(\ref{g31}) for the phase space integration, as well as the treatment of the transitions to $J/\psi$ as two-body decay can and should be refined by using the actual experimental pion spectra, once more detailed data become available. For the purpose of the present discussion we use the described simplifications and illustrate in Fig.~1 the energy behavior of the yield in each decay channel in the suggested model. Clearly, the shape of the curve for the $h_c \pi \pi$ channel does not depend on the mixing angle, and is sensitive only to the widths of the two resonances and the mass splitting between them. In the plots of Fig.~1 the masses of the resonances are fixed at 4.26\,GeV and 4.36\,GeV, and we find that choosing $\Gamma[Y(4260)]=80\,$MeV and $\Gamma[Y(4360)]=100\,$MeV produces a ratio of the production rates at  4.26\,GeV and 4.36\,GeV for the channel $h_c \pi \pi$, that is in a reasonable agreement with the recently reported data~\cite{beslp}. These chosen values of the resonance widths do not exactly coinside with the central values in the Tables~\cite{pdg} ($108 \pm 12\,$MeV and $74 \pm 18\,$MeV), but are compatible with the data, given their present uncertainty. The relative yield at the two peaks in each of the channels $J/\psi \pi \pi$ and $\psi' \pi \pi$ is sensitive to the mixing angle $\theta$, and we find that the value $\theta=40^\circ$ used in the plots appears to not contradict the current data.
\begin{figure}[ht]
\begin{center}
 \leavevmode
    \epsfxsize=13cm
    \epsfbox{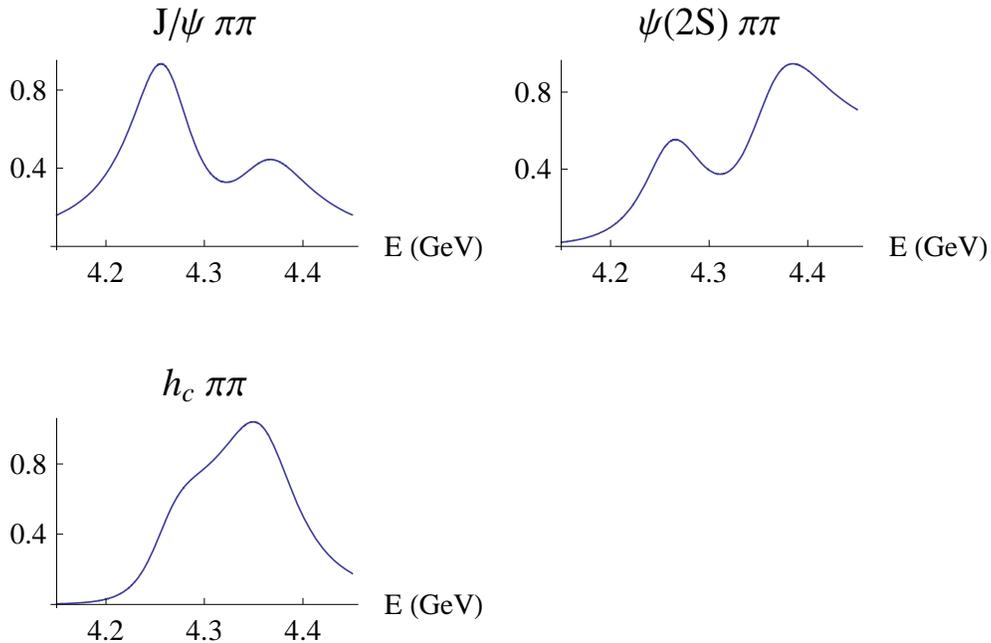}
    \caption{The energy dependence of the yield of the final states $J/\psi \pi \pi$, $\psi' \pi \pi$ and $h_c \pi \pi$ (arbitrary units)  in the region of the $Y(4260)$ and $Y(4360)$ in the discussed model with mixing of two states. }
\end{center}
\end{figure} 

Although we make no attempt here to analyze the relative production rates between different channels, one may notice that the final state $\psi' \pi \pi$ is strongly kinematically suppressed in comparison with $J/\psi \pi \pi$. Therefore, in order to explain the production of $\psi' \pi \pi$ at the $Y(4360)$ peak with a rate comparable to that of $J/\psi \pi \pi$ at $Y(4260)$, the coupling of the state $\Psi_3$ to $\psi' \pi \pi$ should be significantly stronger than to  $J/\psi \pi \pi$. In the hadrocharmonium picture~\cite{hc} this implies that the state $\Psi_3$ contains mostly $\psi'$, rather than $J/\psi$, which is quite natural, given the larger spatial size of the excited charmonium. In other words, the so far observed `affinity' of $Y(4260)$ to $J/\psi$ is a superficial kinematical effect and the underlying hadronic structure in fact contains mostly $\psi'$.

One can also notice the discussed here mixing model predicts a definite pattern of the interference between the resonances. Namely, the phase between the Breit-Wigner factors is $0^\circ$ for the heavy quark spin conserving channels $J/\psi \pi \pi$ and $\psi' \pi \pi$ [Eq.(\ref{a3})], and is $180^\circ$ for the spin violating final state $h_c \pi \pi$ [Eq.(\ref{a1})]. In particular, this phase relation results in the absence of a visible dip in the production rate of the latter final state at energies between the resonances, as can be seen in Fig.~1. 

The large value of the mixing angle, $\theta \approx 40^\circ$, justifies considering the mixing to be the dominant source of the heavy quark spin symmetry breaking and neglecting other possible (smaller) effects of violation of this symmetry, e.g. in the decay amplitudes. Simultaneously the large mixing implies that the unmixed states $\Psi_3$ and $\Psi_1$ are very close in mass. Indeed one can readily solve the two state mixing, and find that at $\theta=40^\circ$ the diagonal masses of $\Psi_3$ and $\Psi_1$ should be approximately 4.30\,GeV and 4.32\,GeV, while the heavy quark spin symmetry breaking mixing amplitude is $\mu \approx 50\,$MeV. The latter amplitude is of a normal scale expected for the spin symmetry violating effects in the charm sector. On the other hand, the proximity of the unmixed states in mass to within about 20\,MeV may appear acidental, but to the best of our knowledge cannot be ruled out. In this context the state $\Psi_3$ can be viewed e.g. as $\psi'$ embedded in a light-quark mesonic excitation with quantum numbers $J^{PC}=0^{++}$, while $\Psi_1$ is an $h_c$ bound in an excited $0^{-+}$ light-quark mesonic state. This possible picture of the unmixed hadrocharmonium states, might require a clarification, regarding the quantum numbers of the light degrees of freedom in the discussed two-pion transitions. For the state $\Psi_3 \sim (1^{--})_{c \bar c} \otimes (0^{++})_{q \bar q}$ the picture of the transition is quite straightforward: both pions in the decay to $J/\psi \pi \pi$, or $\psi' \pi \pi$ can be emitted in the $S$-wave by the $0^{++}$ component, so that no transfer of angular momentum to the $c \bar c$ pair is necessary. The picture is however necessarily different for the decay $\Psi_1 \to h_c \pi \pi$. Indeed, for soft pions the amplitude of the latter decay has the form~\cite{mv86,bgmmv} 
\be
A(\Psi_1 \to h_c \pi \pi) \propto \epsilon_{ijk} \, h_{ci}  \Psi_{1j}  (E_2  p_{1k} + E_1  p_{2k})~,
\label{amph2}
\ee
where $\vec \Psi_1$ and $\vec h_c$ are the polarization amplitudes of the initial $\Psi_1$ and the final $h_c$, and $\vec p_1, \, \vec p_2$ ($E_1, \, E_2$) are the momenta (energies) of the two pions. One can thus see that the angular momentum of the $c \bar c$ pair has to be rotated. This however does not imply a violation of the heavy quark spin symmetry, since it is not the spins of the heavy quarks but rather their ($P$-wave) angular momentum that is rotated. The amplitude (\ref{amph2}) can thus be represented as arising from the action on the initial state of the operator ${\cal O} = \ell_i (E_2  p_{1i} + E_1  p_{2i})$, involving the operator $\vec \ell$ of the angular momentum of the heavy quark pair. Clearly, the operator ${\cal O}$ has the quantum numbers $0^{-+}$, so that the simplest hadrocharmonium configuration, linked by this operator to the final state $h_c \pi \pi$, is  $\Psi_1 \sim (1^{+-})_{c \bar c} \otimes (0^{-+})_{q \bar q}$~\footnote{One can also notice that the quantum numbers of the emitted dipion in its center of mass frame are $0^{++}$ and $2^{++}$, and these combine with the angular momentum in the rest frame of the heavy quarkonium to ensure the conservation of the overall angular momentum and the parity.}. 

It can be noted that the possible latter structure of the state $\Psi_1$ also suggests that there can be a substantial yield in the not yet observed channel $e^+e^- \to h_c \eta$ in the same energy range of the $Y(4260)$ and $Y(4360)$ resonances. Due to the present uncertainty in understanding the conversion of the light degrees of freedom in hadrocharmonium into light mesons it is difficult to offer a specific prediction for the cross section. It is quite possible however that the yield of the $h_c \eta$ final state can be comparable to that of $h_c \pi \pi$.

Furthermore, as previously discussed~\cite{hc,dgv}, combining the light-matter excitations with the states of charmonium generally gives rise to a number of new charmonium-like resonances. In particular, in the discussed here picture, besides the states $\Psi_3 \sim (1^{--})_{c \bar c} \otimes (0^{++})_{q \bar q}$ and $\Psi_1 \sim (1^{+-})_{c \bar c} \otimes (0^{-+})_{q \bar q}$ one might expect existence of hadrocharmonium states with the structure $(1^{--})_{c \bar c} \otimes (0^{-+})_{q \bar q}$ (with a mass approximately 3.9\,GeV), and $(1^{+-})_{c \bar c} \otimes (0^{++})_{q \bar q}$ (at approximately 4.7\,GeV). It is clear however that these isoscalar resonances should have quantum numbers $J^P=1^{+-}$ and would not be directly produced in $e^+e^-$ annihilation, or in single pion transitions from the states produced in $e^+e^-$ annihilation.

In summary. We suggest that the recently reported significant violation of the heavy quark spin symmetry at the $Y(4260)$ and $Y(4360)$ peaks is due to these resonances being mixtures of hadrocharmonium states containing spin triplet and spin singlet charmonium. Although we cannot explain the accidental proximity in mass of the two unmixed hadrocharmonium states, the suggested picture appears to be in agreement with the currently known data. In particular the energy dependence of the yield of final states is sensitive to the parameters of the mixing, so that a detailed study of this behavior can put the suggested scheme to a further experimental test. Furthermore, the suggested mixing scheme predicts a distinctive pattern of interference between the resonances which also can be studied experimentally.

{\it Note added.} Very shortly after the initial version of this paper was placed in the arXive, the BESIII experiment made available~\cite{beshc} the data on the energy dependence of the cross section for the process $e^+e^- \to h_c \pi^+ \pi^-$ in the range of $E$ from 4.19\,GeV to 4.42\,GeV. We thus report here on our attempt at fitting the data within the suggested two resonance model, with the interference between the resonances described by Eq.(\ref{a1}). In performing the fit we allowed the masses and the widths in the two Breit-Wigner factors to float, as well as the overall normalization factor, thus resulting in the total of five fit parameters. Furthermore we included only the statistical experimental errors in our calculation of $\chi^2$, and not included the reported systematical errors. We believe that this is the proper procedure, since the systematical errors in the data~\cite{beshc} arise from the uncertainty in the normalization and are in fact proportional to the central values of the data at each energy point, thus being strongly correlated. Since the overall normalization is one of our fit parameters, the experimental uncertainty in the normalization is absorbed into the definition of this overall factor~\footnote{We thank A.Bondar for an illuminating discussion of this point.}. It should be also noted that, as discussed in the paper, using the low energy approximation for the pion emission amplitude introduces an uncertainty at higher energies. A better procedure would require the data on the actual pion spectra at each energy. Lacking such data, we estimate the effect of the possible inaccuracy of our treatment of the higher energies by comparing the results of the fit to all the available data and to the same data with the highest energy point excluded.  We find that the extracted parameters of the lower mass resonance and the overall quality of the fit are quite stable under this variation of the procedure, while the most affected is the width $\Gamma_2$ of the heavier resonance. Also with the limited experimental information available to us at present, we do not attempt to evaluate the errors in the extracted parameters and quote here only the `central' values corresponding to the minimum of $\chi^2$.
The result of our fit, using all ten data points from 4.19\,GeV to 4.42\,GeV, for the masses and the widths of the resonances is $M_1=4213\,$MeV, $\Gamma_1=69\,$MeV, $M_2=4379\,$MeV, $\Gamma_2 = 160\,$MeV with $\chi^2/N= 6.0/5$, where $N$ is the number of degrees of freedom. The fit with the data point at $E = 4.42\,$GeV excluded yields $M_1=4214\,$MeV, $\Gamma_1=61\,$MeV, $M_2=4351\,$MeV, $\Gamma_2 = 117\,$MeV with $\chi^2/N= 3.6/4$. The input data and the fit curves are shown in Fig.2.
\begin{figure}[ht]
\begin{center}
 \leavevmode
    \epsfxsize=13cm
    \epsfbox{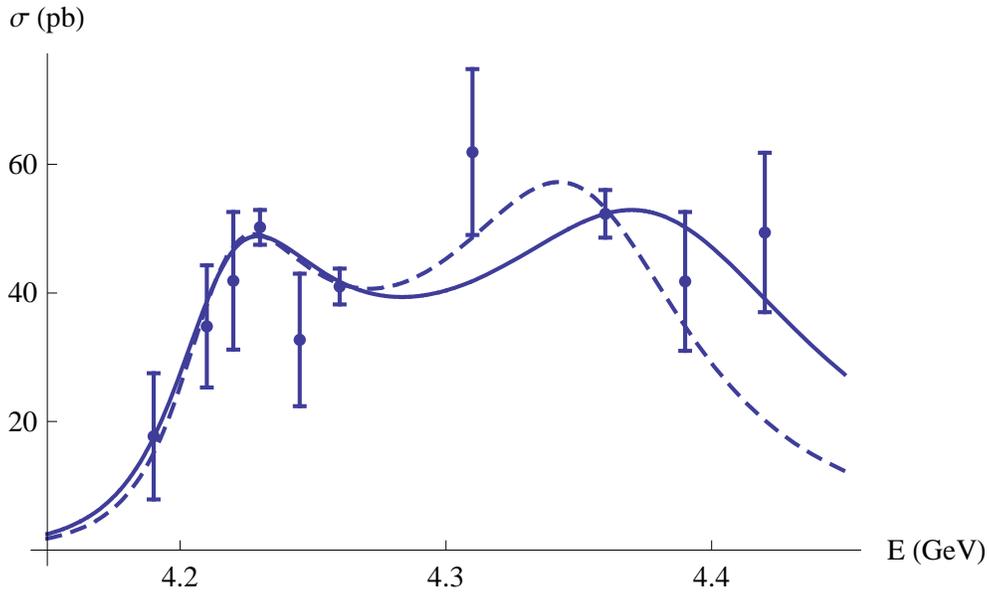}
    \caption{The energy dependence of the cross section $\sigma(e^+e^- \to h_c \pi^+ \pi^-)$. The data points from Ref.~\cite{beshc} are shown with the statistical errors only. The curves show the behavior in the two-resonance model with the parameters determined from the fit with all the data points (solid) and with the highest energy data point excluded (dashed). }
\end{center}
\end{figure} 

One can readily notice that our fit results in a lower, than the table value, mass of the lower resonance $Y(4260)$, $M_1 \approx 4215\,$MeV. This low value is compatible within the errors with the one reported in Ref.~\cite{belle26}, but does not appear to agree with Refs.~\cite{babar26,cleo26,babar26:12}. It should be noted however that all the previous determinations were done using the final state $J/\psi \pi \pi$. The production of this final state, as well as of $\psi' \pi \pi$ requires no violation of the heavy quark spin symmetry and may receive an unsuppressed contribution from the non resonant continuum. The interference between the $Y(4260)$ resonance and the continuum amplitude may generally result in a shift of the apparent position of the resonance. It is not clear however whether this shift can be large enough to explain the difference between the results of our fit to the data~\cite{beshc} and the previous determinations. In either case the significance of this discrepancy can possibly be understood with a more detailed set of data.
 
This work is supported, in part, by the DOE grant DE-FG02-94ER40823.

\end{document}